# Evidence for a Vestigial Nematic State in the Cuprate Pseudogap Phase

Sourin Mukhopadhyay[1,2], Rahul Sharma[2,3], Chung Koo Kim[3], Stephen D. Edkins[4], Mohammad H. Hamidian[5], Hiroshi Eisaki[6], Shin-ichi Uchida[6,7], Eun-Ah Kim[2], Michael J. Lawler[2], Andrew P. Mackenzie[8], J. C. Séamus Davis[9,10] and Kazuhiro Fujita[3]

1. *Department of Physics, IIST, Thiruvananthapuram 695547, India*
2. *LASSP, Department of Physics, Cornell University, Ithaca, NY 14853, USA*
3. *CMPMS Department, Brookhaven National Laboratory, Upton, NY 11973, USA*
4. *Department of Applied Physics, Stanford University, Stanford, CA 94305, USA*
5. *Department of Physics, Harvard University, Cambridge, MA, USA*
6. *Institute of Advanced Industrial Science and Technology, Tsukuba, Ibaraki 305-8568, Japan*
7. *Dept. of Physics, University of Tokyo, Bunkyo-ku, Tokyo 113-0033, Japan*
8. *Max Planck Institute for Chemical Physics of Solids, D-01187 Dresden, Germany*
9. *Department of Physics, University College Cork, Cork, T12 R5C, Ireland*
10. *Clarendon Laboratory, University of Oxford, Parks Rd., Oxford, OX1 3PU, UK*

**The $CuO_2$ antiferromagnetic insulator is transformed by hole-doping into an exotic quantum fluid usually referred to as the pseudogap (PG) phase. Its defining characteristic is a strong suppression of the electronic density-of-states $D(E)$ for energies $|E|<\Delta^*$, where $\Delta^*$ is the pseudogap energy. Unanticipated broken-symmetry phases have been detected by a wide variety of techniques in the PG regime, most significantly a finite $Q$ density-wave (DW) state and a $Q=0$ nematic (NE) state. Sublattice-phase-resolved imaging of electronic structure allows the doping and energy dependence of these distinct broken symmetry states to be visualized simultaneously. Using this approach, we show that, even though their reported ordering temperatures $T_{DW}$ and $T_{NE}$ are unrelated to each other, both the DW and NE states always exhibit their maximum spectral intensity at the same energy, and using independent measurements that this is the pseudogap energy $\Delta^*$. Moreover, no new energy-gap opening coincides with the appearance of the DW state (which should theoretically open an energy gap on the Fermi-surface), while the observed** **pseudogap opening coincides with the appearance of the NE state (which should theoretically be incapable of opening a Fermi-surface gap). We demonstrate how this perplexing phenomenology of thermal transitions and energy-gap opening at the breaking of two highly distinct symmetries can be understood as the natural consequence of a vestigial nematic state within the pseudogap phase of $Bi_2Sr_2CaCu_2O_8$.**

*Keywords:* Cuprate, Pseudogap, Broken Symmetry, Density Wave, Vestigial Nematic

**Significance Statement:**

In the cuprate pseudogap phase, an energy gap of unknown mechanism opens, and both an electronic nematic phase (NE) and a density-wave phase (DW) appear. Perplexingly, the DW which should generate an energy gap, appears without any new gap opening; and the NE which should be incapable of opening an energy gap, emerges coincident with the pseudogap opening. Recently, however, Nie *et al*. demonstrated theoretically that a disordered unidirectional DW can generate a vestigial nematic (VN) phase. If the cuprate pseudogap phase were in such a VN state, the energy gap of the NE and DW should be identical to each other and to the pseudogap. We report discovery of such a phenomenology throughout the phase diagram of underdoped $Bi_2Sr_2CaCu_2O_8$.

*Symmetry Breaking in the Cuprate Pseudogap Regime*

The cuprate pseudogap energy $\Delta^*$ is defined to be the energy range within which significant $D(E)$ suppression occurs in underdoped cuprates[1,2]. Figure 1a shows measured $D(E)$ for hole-density $p$=0.06 in $Bi_2Sr_2CaCu_2O_{8+\delta}$, with arrow indicating a typical example of the feature in $D(E)$ that occurs at $E=\Delta^*$. The inset shows the measured evolution of $D(E)$ and this signature of $\Delta^*$, with $p$. Figure 1b shows the consequent doping-dependence of $\Delta^*(p)$ for this sequence of $Bi_2Sr_2CaCu_2O_{8+\delta}$ samples. This exemplifies the well-known correspondence[1] between $\Delta^*(p)$ and the temperature $T^*(p)$ at which the pseudogap phenomenology appears; both fall linearly with increasing $p$ until reaching zero at approximately $p \approx 0.19$. In the last decade, extensive evidence has emerged for electronic symmetry breaking at or below $T^*(p)$ as shown schematically in Fig. 1c. For $T < T^*(p)$ and $p < p_C \approx 0.19$, two distinct classes of broken-symmetry states are widely reported. The first is a nematic state; it occurs at wavevector $\boldsymbol{Q}=0$ and generally exhibits breaking of 90º-rotational (C$_4$) symmetry[3-16]; sometimes additional $\boldsymbol{Q} = 0$ broken symmetries are detected depending on the experimental technique. These phenomena have been reported using multiple techniques in YBa$_2$Cu$_3$O$_{7-x}$, Bi$_2$Sr$_2$CaCu$_2$O$_{8+x}$, Bi$_2$Sr$_2$CuO$_{6+x}$ and HgBa$_2$CuO$_{4+\delta}$, and are observed to occur below a characteristic temperature $T_{NE}(p)$, which appears to fall linearly with increasing $p$ along the trajectory shown in Fig. 1c. Already this is a conundrum: while the onset temperature $T_{NE}(p)$ of this broken symmetry tracks the temperature $T^*(p)$ at which pseudogap opens (Fig. 1c), within mean-field theory of electron fluids ordering at $\boldsymbol{Q}=0$ it cannot open an energy-gap on the Fermi surface. The second ordered state within the pseudogap phase breaks translational symmetry with axial wavevectors $\boldsymbol{Q} = (Q,0); (0,Q)$ parallel to the CuO$_2$ axes[17-27]. (Here we do not discuss the DW state generated by very high magnetic fields[26].) This zero-field DW state has been reported to occur in La$_{2-x}$Ba$_x$CuO$_4$, YBa$_2$Cu$_3$O$_{7-x}$, Bi$_2$Sr$_2$CaCu$_2$O$_{8+x}$, Bi$_2$Sr$_2$CuO$_{6+x}$, HgBa$_2$CuO$_{4+x}$, and Ca$_{2-x}$Na$_x$CuO$_2$Cl$_2$. In YBa$_2$Cu$_3$O$_{7-x}$, Bi$_2$Sr$_2$CuO$_{6+x}$, Bi$_2$Sr$_2$CaCu$_2$O$_{8+x}$ and Ca$_{2-x}$Na$_x$CuO$_2$Cl$_2$ samples, the DW exhibits a predominantly $d$-symmetry form factor[28-32]. The DW phenomenology reveals a second puzzle: although a DW with finite $\boldsymbol{Q}$ should open a mean-field energy-gap spanning the Fermi energy (while folding the band-structure due to the larger electronic unit-cell), no energy-gap is reported to open specifically at the cuprate DW onset temperature $T_{DW}(p)$. Moreover, $T_{DW}(p)$ is manifestly unrelated to the temperature $T^*(p)$ at which the pseudogap opens (Fig. 1c). Thus, the essence of the broken-symmetry riddle of cuprates is that the DW state that could produce an energy-gap in $D(E)$ is only detectable at temperatures $T_{DW}(p) << T^*(p)$ and without the opening of a new energy-gap, while the NE state that appears at the temperature $T_{NE}(p) \approx T^*(p)$ where the pseudogap opens should be incapable of opening a Fermi surface energy-gap. No comprehensive explanation has been proposed to address these linked mysteries.

Both types of broken-symmetry states can be studied simultaneously in the same field of view (FOV) using high resolution spectroscopic imaging scanning tunneling microscopy (SISTM) [33]. To do so, Fourier analysis of sublattice phase-resolved electronic structure imaging of the $CuO_2$ plane is required[8,28,29,30]. The tunnel-current $I(r,E)$ and tip-sample differential tunneling conductance $\frac{dI}{dV}(r, E = eV) \equiv g(r,E)$ are first measured at bias voltage $V=E/e$ and with spatial resolution of typically 7X7 pixels within each $CuO_2$ unit cell. To suppress serious systematic "set-point" errors[33] we evaluate $Z(r,E) = g(r,E)/g(r,-E)$, a function for which distances, wavelengths, spatial-phases and energy-magnitudes of electronic structure can be measured accurately. The necessary spatial-phase accuracy in $g(r,E)$ and $Z(r,E)$ is achieved using a picometer-scale transformation that renders the topographic image $T(\mathbf{r})$ perfectly $a_0$-periodic, and is then applied to $g(r,E)$ to register all the electronic-structure data to the ideal $CuO_2$ lattice[8,28,29,30]. The necessity of these advanced SI-STM techniques to measure the cuprate broken symmetry phases can be seen directly in Fig. 2a which shows a typical $Z(r,E)$, with local breaking of $C_4$ symmetry within the $CuO_2$ cell (NE) and its periodic modulations (DW); a schematic overlay is used to identify the locations of Cu and O orbitals. Figure 2b is the Fourier transform of a larger-FOV $Z(r, E\sim100meV)$ image of $Bi_2Sr_2CaCu_2O_{8+x}$ at $p\sim0.06$ containing the two basic signatures of electronic symmetry breaking, the Bragg peaks at $\mathbf{Q}^B = [(1,0)(0,1)]2\pi/a_0$ and the peaks near $\mathbf{Q}^D \approx ((0.75,0)(0,0.75))2\pi/a_0$ due to the $d$-symmetry form factor DW[28,29,30].

The $\boldsymbol{Q}$=0 rotational symmetry-breaking state (NE) is detected by resolving both the real $ReZ(\boldsymbol{Q^B}, E)$ and imaginary $ImZ(\boldsymbol{Q^B}, E)$ components of the Bragg amplitudes of $Z(\boldsymbol{q}, E)$, to form a direct measure of intra-unit cell nematicity[8]

$$N^Z(E) \equiv ReZ(\boldsymbol{Q}^B_y, E) - ReZ(\boldsymbol{Q}^B_x, E) \qquad (1)$$

where $\boldsymbol{Q}^B_y$ and $\boldsymbol{Q}^B_x$ are the two Bragg wavevectors along the $y$-axis/$x$-axis respectively. Any spurious effects of STM tip anisotropy on $N^Z(E)$ are ruled out as described in SI Appendix Section 1. If the measured $N^Z(E)$ is then non-zero, breaking of C$_4$-symmetry is occurring due to inequivalence on the average, of electronic structure at the two oxygen sites within each CuO$_2$ cell[8,28,29,30,31,32,33]. Figure 3a-e shows the measured energy dependence of $ReZ(\boldsymbol{Q}^B_y, E)$, $ReZ(\boldsymbol{Q}^B_x, E)$ and their difference $N^Z(E)$ for Bi$_2$Sr$_2$CaCu$_2$O$_{8+x}$ at $p$=0.06, 0.08, 0.10, 0.14 and 0.17. Figure 3f then summarizes the doping dependence of the energy, $E^N_{max}$, at which $N^Z(E)$ achieves its maximum intensity. Thus, the characteristic energy of maximum spectral intensity of this measure of cuprate nematicity diminishes approximately linearly with increasing $p$.

The translational symmetry breaking (DW) state exhibits a predominantly $d$-symmetry form factor ($d$FF) because the DW is based on modulating the inequivalence between the $O_x$ and $O_y$ sites within each CuO$_2$ unit-cell[8,28,29,30,31,32,33] at wavevector $\boldsymbol{Q}^D_x$, as discussed in SI Appendix Section 2. Such modulations are described by $A(\boldsymbol{r}) = D(\boldsymbol{r})Cos(\phi(\boldsymbol{r}) + \phi_0(\boldsymbol{r}))$, where $\phi(\boldsymbol{r}) = \boldsymbol{Q}^D_x \cdot \boldsymbol{r}$ is the DW spatial phase at location $\boldsymbol{r}$ and $D(\boldsymbol{r})$ is the magnitude of the $d$-symmetry form factor. $D(\boldsymbol{r}) = A_D$ when $r = r_{O_x}$, $D(\boldsymbol{r}) = -A_D$ when $r = r_{O_y}$ and $D(\boldsymbol{r}) = 0$ otherwise, where $A_D$ is a real number for the amplitude. To study this state, the sublattice-phase-resolved $Z(\boldsymbol{r}, E)$ images are separated into three: $Cu(\boldsymbol{r})$, containing only Cu sites and, $O_x(\boldsymbol{r})$; $O_y(\boldsymbol{r})$, containing only the $x$/$y$-axis oxygen sites. Phase-resolved Fourier transforms of the $O_x(\boldsymbol{r},E)$ and $O_y(\boldsymbol{r},E)$ sublattice images, $\tilde{O}_x(\boldsymbol{q}, E) = Re\tilde{O}_x(\boldsymbol{q}, E) + iIm\tilde{O}_x(\boldsymbol{q}, E)$; $\tilde{O}_y(\boldsymbol{q}, E) = Re\tilde{O}_y(\boldsymbol{q}, E) + iIm\tilde{O}_y(\boldsymbol{q}, E)$, are then used to determine the form factor symmetry for modulations at any $\boldsymbol{q}$

$$D^Z(\boldsymbol{q}, E) = (\tilde{O}_x(\boldsymbol{q}, E) - \tilde{O}_y(\boldsymbol{q}, E)) \quad (2a)$$

$$S'^Z(\boldsymbol{q}, E) = (\tilde{O}_x(\boldsymbol{q}, E) + \tilde{O}_y(\boldsymbol{q}, E)) \quad (2b)$$

Figure 4a-e show our typical images of $Z(\boldsymbol{r}, E \sim \Delta^*)$ while Fig. 4f-j show $|D^Z(\boldsymbol{q}, E \sim \Delta^*)|^2$ the *d*-symmetry form factor, power spectral density Fourier transforms derived from 4a-e. The peaks identified by red dashed circles are the signature of the *d*-symmetry form factor DW state[28,29,30,31,32]. At atomic scale, these features in $D^Z(\boldsymbol{q}, E \sim \Delta^*)$ represent two populations of nanoscale unidirectional *d*-symmetry form factor DW domains[29].

### *Evolution of Broken Symmetry Spectral-Weight with Hole Density*

Our second key objective is then to measure the doping dependence of the energy, $E_{max}^D$, at which the DW order $D^Z(\boldsymbol{Q}^D, E)$ achieves its maximum spectral intensity. Figure 5a shows the measured $|D^Z(\boldsymbol{Q}^D, E)|$ integrated within the dashed circles in Fig. 4f-j for *p*=0.6, 0.8, 0.10, 0.14 and 0.17 plotted versus energy *E*. The *d*-symmetry form factor is negligible for modulations in the low energy range of Bogoliubov quasiparticles (Figs 4k-o), but $D^Z(\boldsymbol{Q}^D, E)$ rapidly becomes intense at higher energy before finally diminishing quickly at highest *E*. Evidently, for each *p* there is a specific energy $E_{max}^D$ at which a maximum in $|D^Z(\boldsymbol{Q}^D, E)|$ occurs. The values of $E_{max}^D$ are indicated by the black arrows in Fig. 5a, and these we associate with the energy gaps of the disordered DW (SI Appendix Section 3). Figure 5b then summarizes the doping dependence of measured $E_{max}^D$, revealing that that the characteristic energy of maximum spectral intensity of the DW state also diminishes approximately linearly with increasing *p*.

To appreciate the implications of these observations, it is important to realize that the order parameters $N^Z(E)$ in Eqn. 1 and $D^Z(\boldsymbol{q}, E)$ in Eqn. 2a use no common *q*-space data whatsoever, despite having been measured in the identical real-space field of view. The key experimental result of these studies is then shown in Figure 5b: a combined plot of measured $E_{max}^N(p)$, $E_{max}^D(p)$ versus the measured values of *Δ*\*(*p*) for underdoped

Bi$_2$Sr$_2$CaCu$_2$O$_{8+x}$ samples with $0.06 \leq p \leq 0.17$. The striking fact it reveals is that the two characteristic energies $E_{max}^N$ and $E_{max}^D$, where the distinct broken-symmetry states NE and DW have maximal spectral-intensity, are indistinguishable from each other throughout this wide region of the underdoped phase diagram. Even more notably, both $E_{max}^N(p)$ and $E_{max}^D(p)$ follow a trajectory with doping that is indistinguishable within error bars from that of the independently measured pseudogap energy $\Delta^*(p)$ (Fig. 1a,b). Therefore, we find that the characteristic energy of maximal spectral intensity for both cuprate broken-symmetry states is always the pseudogap energy $\Delta^*(p)$. These results add yet another conundrum to our list: how could the characteristic energies of NE and DW states be virtually identical to each other and to the PG energy $\Delta^*$ (Fig. 5b) over this wide range of doping, while the onset temperatures $T_{NE}(p)$ and $T_{DW}(p)$ of these phases appear unrelated to each other and to vary quite differently with doping (Fig. 1c)? Combined with the question of why the DW state (which should generate an energy gap) appears without any new gap opening at $T_{DW}(p)$, while the NE state (which should be incapable of opening an energy gap) emerges at the pseudogap opening temperature $T_{NE}(p) \approx T^*(p)$ (Fig. 1c), this presents an intricate puzzle.

## *Analysis*

In pursuit of its resolution, we note that electronic liquid crystal states of a doped Mott insulator are naturally expected to exhibit distinct NE and DW broken-symmetry phases[34,35,36]. Within this context, the critical temperatures and characteristic electronic energies of these two highly distinct phases are not required to be related, except for the fact that the nematic phase should appear first with falling temperature. Recently, however, the effects of quenched disorder on the quasi-two-dimensional version of this situation has been discovered[37,38]. The remarkable conclusion is that, while long range order of a unidirectional incommensurate finite-*Q* DW cannot exist in the presence of quenched disorder, its short-range relict survives up to a certain critical disorder strength but in the form of a ***Q***=0 broken rotational-symmetry state. This state was dubbed a *vestigial nematic*[37,38] (VN). Note that, as with the unidirectional *s*-symmetry form factor DW state considered in Ref. 37, a unidirectional *d*-symmetry form factor DW state (as in underdoped cuprates) also breaks local translational and rotational symmetries[28,29,30], and should also yield a VN state in the presence of strong disorder. Several distinguishing consequences are predicted within such VN theories. First, the energy scale of the NE state for which rotational symmetry is broken globally should always be the same as that of the DW state in which translational symmetry is broken locally, since the two effects stem from the identical microscopic electronic phenomenon (unidirectional DW with disorder). Second, as temperature diminishes below the phase transition where the vestigial nematic state appears and its gap opens, the intensity of the short-range DW underpinning the VN should increase continuously as thermal fluctuations are reduced but with no new phase transition and no additional gap opening, as indicated schematically by dashed contours in Fig. 5c. Here, the horizontal axis is carrier density but, because strong spatial correlations exist between dopant atom distributions and electronic disorder in Bi$_2$Sr$_2$CaCu$_2$O$_{8+x}$[39], it also corresponds with disorder strength[37,38] for that material.

Vestigial nematic theories cast a fresh light on the perplexing phenomenology of the NE[3-16] and DW[17-27] symmetry-breaking in the cuprate pseudogap phase (Figs 2-5). If this phase is a VN state, then a highly disordered DW opens a gap $\Delta^*$ at T* but without long-range DW order, so that only a vestigial nematic phase appears. The VN model then implies that all three energies: the pseudogap, the characteristic energy gap of the DW, and the apparent energy gap of the NE state should all coincide at every doping. This is because all three signatures derive from the same microscopic phenomenon: the appearance of the unidirectional but disordered DW state. Coincidence of measured $E_{max}^N(p)$, $E_{max}^D(p)$ and $\Delta^*(p)$ is precisely what we report in Fig. 5b, for underdoped Bi$_2$Sr$_2$CaCu$_2$O$_{8+x}$ samples over a wide doping range $0.06 \leq p \leq 0.17$. Moreover, in a VN state with phase transition at $T_{NE}(p)$, the onset temperatures $T_{NE}(p)$ and $T_{DW}(p)$ should be unrelated to each other and vary independently with doping, provided $T_{DW}(p)$ is an instrumentation-limited threshold for detection of the disordered charge modulations rather than a phase transition. In that case, detection of DW phenomena should not coincide with opening of a new gap at $T_{DW}(p)$ because the DW gap has already opened at $T_{NE}(p) \approx T^*(p)$ when the VN phase transition occurs. Thus, VN theory[37,38] appears highly consistent with the experimental observations reported herein (Fig. 5b), and more generally with the intricate phenomenology of thermal transitions, energy-gap opening, and distinct symmetry-breaking in underdoped cuprates[3-32] (Fig. 1c). We note, however, that neither VN theory nor our studies yet discriminate whether the fundamental DW state with wavevector $\boldsymbol{Q}$ is a conventional charge density wave with order parameter $\langle c_k^\dagger c_{k+Q} \rangle$ or is a pair density wave[40] with order parameter $\langle c_k^\dagger c_{-k+Q}^\dagger \rangle$.

**Acknowledgements** S.M. acknowledges support from NSF Grant DMR- 1719875 to the Cornell Center for Materials Research; S.U. and H.E. acknowledge support from a Grant-in-Aid for Scientific Research from the Ministry of Science and Education (Japan) and the Global Centers of Excellence Program for Japan Society for the Promotion of Science; K.F. and C.-K. Kim acknowledge salary support from the US Department of Energy, Office of Basic Energy Sciences, under contract number DEAC02-98CH10886. E.-A.K. acknowledges Simons Fellow in Theoretical Physics Award #392182 and from the from the US Department of Energy, Office of Basic Energy Sciences, under grant number DE-SC0018946.; SDE acknowledges studentship funding from the EPSRC under Grant EP/G03673X/1; M.H.H, S.D.E and J.C.S.D acknowledge support from the Gordon and Betty Moore Foundation's EPiQS Initiative through Grant GBMF4544; J.C.S.D. acknowledges support from Science Foundation Ireland under Award SFI 17/RP/5445 and from the European Research Council (ERC) under Award DLV-788932.

**Author Contributions:** S.M., C.K.K, M.H.H., S.D.E., and K.F. carried out the experiments; S.M., C.K.K, R.S. and K.F. carried out the analysis; E.A.K and M.J. L provided theoretical motivation and guidance; H.E. and S.U. synthesized the samples, J.C.D. A.P.M. and K.F. supervised the project and wrote the paper with key contributions from S.M., R.S., C.K.K, M.H.H., and S.D.E. The manuscript reflects the contributions and ideas of all authors.

**Declaration:** H. Eisaki and R. Comin are co-authors on a 2015 article "*Symmetry of charge order in cuprates*", *Nature Materials* **14**, 796-800 (2015)

# FIGURE CAPTIONS

**Figure 1: Doping Evolution of Cuprate Symmetry Breaking and Pseudogap**

**a.** Measured D(E) at p=0.06 and T=2.2K in $Bi_2Sr_2CaCu_2O_{8+\delta}$. This $dI/dV(E = eV) \equiv g(E)$ curve is a typical tunneling spectrum of underdoped cuprates with empty-state DOS-peak at energy $\Delta^*$. Inset shows the characteristic $g(E)$ for the range of $Bi_2Sr_2CaCu_2O_{8+\delta}$ samples (p=0.06, 0.08, 0.10, 0.14 and 0.17) studied. The evolution of $\Delta^*(p)$ is indicated by the black arrows.

**b.** Doping-dependence of the measured average value of ∆∗ (p=0.06, 0.08, 0.10, 0.14 and 0.17, respectively) from the set of $Bi_2Sr_2CaCu_2O_{8+\delta}$ samples studied.

**c.** Colored background represents schematic phase diagram of hole-doped $CuO_2$. Two classes of broken-symmetry states have been widely reported therein. The first indicated by circles occurs at wavevector $Q = 0$ and typically exhibits breaking of $C_4$-symmetry within the $CuO_2$ unit cell that appears at the PG temperature scale T*. The second state indicated by diamonds breaks translational symmetry with wavevector $Q^D = (Q,0); (0,Q)$ exhibiting a predominantly d-symmetry form factor, and often appears at a temperature scale far lower than T*.

**Figure 2: Q=0 and Q≠0 broken-symmetry states imaged simultaneously**

**a.** Real-space phase resolved SI-STM image of d-symmetry form factor density wave (dFF DW) modulations in $Z(r, E = 94meV)$ of a $Bi_2Sr_2CaCu_2O_{8+\delta}$ sample with p=0.06 measured at T=2.2K. Transparent overlay shows the correspondences between the schematic dFF DW and the real-space phase resolved electronic structure. Intra-unit cell $C_4$-symmetry is broken in virtually every $CuO_2$ unit cell, and it is the modulation of this electronic inequivalence between $O_x$ and $O_y$ sites that generates the dFF DW state.

**b.** Typical Fourier transform image $Z(q, E)$ from underdoped $Bi_2Sr_2CaCu_2O_{8+\delta}$. Locations of the Bragg peaks $Q_x^B, Q_y^B$ are indicated by circles.

**Figure 3**: **Doping- and Energy-Evolution of the Q=0 Broken-Symmetry (NE) State**

a. Inset: Measured $ReZ(q, E = 72meV)$ for Bi$_2$Sr$_2$CaCu$_2$O$_{8+\delta}$ with *p*=0.06 at T=2.2K. The magnitude of $N^Z(E) \equiv ReZ(\boldsymbol{Q}_y^B, E) - ReZ(\boldsymbol{Q}_x^B, E)$ is measured as the difference in real intensity of the two Bragg peaks at $\boldsymbol{Q}_x^B, \boldsymbol{Q}_y^B$ (indicated by diamond and star shapes).

a.–e. Measured $ReZ(\boldsymbol{Q}_y^B, E)$, $ReZ(\boldsymbol{Q}_x^B, E)$ and $N(E)$ versus hole-density (*p*=0.06, 0.08, 0.10, 0.14 and 0.17 respectively) in the Bi$_2$Sr$_2$CaCu$_2$O$_{8+\delta}$ samples at T=2.2K reported here. For each $N^Z(E)$ curve, an energy $E_{max}^N$ where $N^Z(E)$ achieves its maximum is indicated by a vertical arrow.

f. Plot of measured $E_{max}^N(p)$ for *p*=0.06, 0.08, 0.10, 0.14 and 0.17. This shows how the energy associated with maximum spectral intensity of the NE state diminishes approximately linearly with increasing *p*.

**Figure 4**: **Doping-Evolution of the Q≠0 Broken-Symmetry (DW) State**

a.–e. Real-space map of phase resolved DW modulations in $Z(r, E)$ measured at T=2.2K for doping *p*=0.06, 0.08, 0.10, 0.14 and 0.17 at E=84, 72, 56, 52 and 46meV, respectively.

f.-j. Corresponding q-space DW power-spectral-densities (PSD) $|D^Z(\boldsymbol{q}, E)|^2$ from Eqn. 2 for dopings p=0.06, 0.08, 0.10 and 0.14 and 0.17, respectively. Red circles represent the relevant regions of interest for d-symmetry form factor DW at wavevectors $\boldsymbol{Q}^D \cong ((0.25,0)(0,0.25))2\pi/a_0$. Locations of the Bragg peaks $Q_x^B, Q_y^B$ are indicated by "+" sign.

k.–o. The power-spectral-densities of Bogoliubov quasiparticle $|S'^Z(\boldsymbol{q}, E)|^2$ from Eqn. 2 for dopings *p*=0.06, 0.08, 0.10 and 0.14 and 0.17, respectively, measured at lower energy on the same sample set. Locations of the Bragg peaks $Q_x^B, Q_y^B$ are indicated by "+" sign.

**Figure 5: Relationship of Characteristic Energies of NE, DW States and the PG**

**a.** Measured energy variation of normalized amplitude for d-symmetry form factor DW versus doping. The corresponding Gaussian fit to the data, used to extract the peak positions, are shown in solid curves. The dFF DW amplitude is extracted by integrating $|D^Z(\boldsymbol{q},E)|^2$ over the primary DW wavevectors at $\boldsymbol{Q}^D \cong ((0.25,0)(0,0.25))2\pi/a_0$ (red circles in **4f-j**). The data for hole-density $p$=0.06, 0.08, 0.10, 0.14 and 0.17 in the Bi$_2$Sr$_2$CaCu$_2$O$_{8+\delta}$ samples are reported here. For each $|D^Z(\boldsymbol{q},E)|^2$ curve, an energy $E_{max}^D$ for maximum spectral weight is indicated by a vertical arrow.

**b.** Plot of measured $E_{max}^N(p)$ and $E_{max}^D(p)$ versus $\Delta^*(p)$ for values of $p$=0.06, 0.08, 0.10, 0.14 and 0.17. This shows how the energy for maximum spectral intensity of both the NE and the DW states diminishes approximately linearly with increasing $p$, and that they are always virtually identical to each other and to the PG energy. Dashed blue line is a guide to the eye.

**c.** A schematic diagram of the hypothesized VN state within the cuprate phase diagram[37] The VN hypothesis appears consistent with all the data and analysis presented throughout this paper. Note that the antiferromagnetic Mott insulator and superconducting regions shown here are not included in the analysis of Ref.37,38

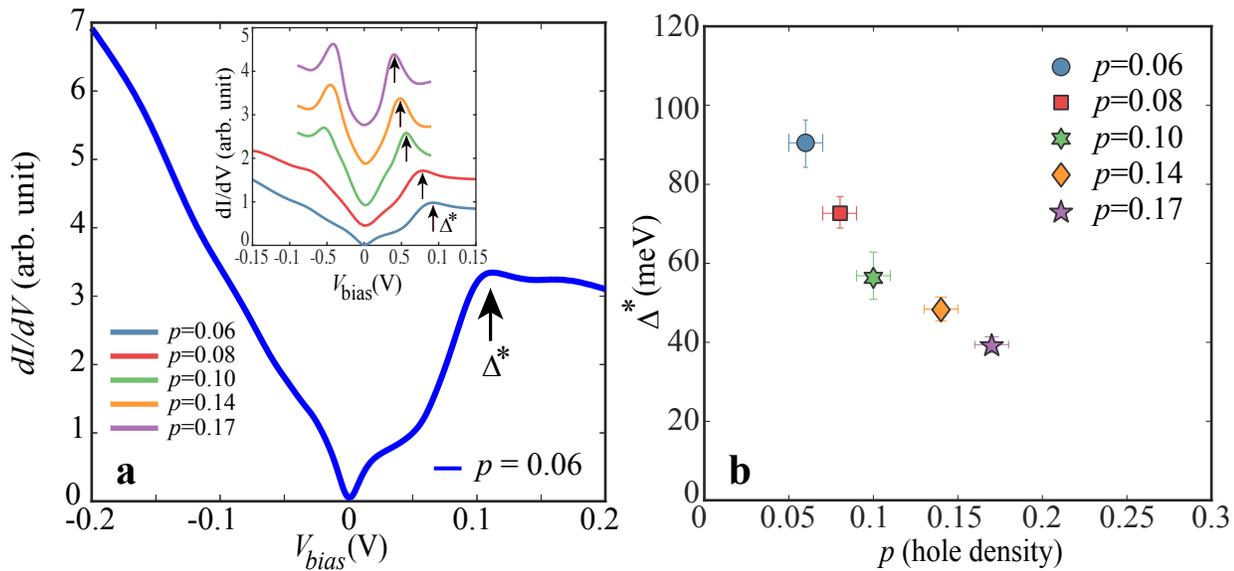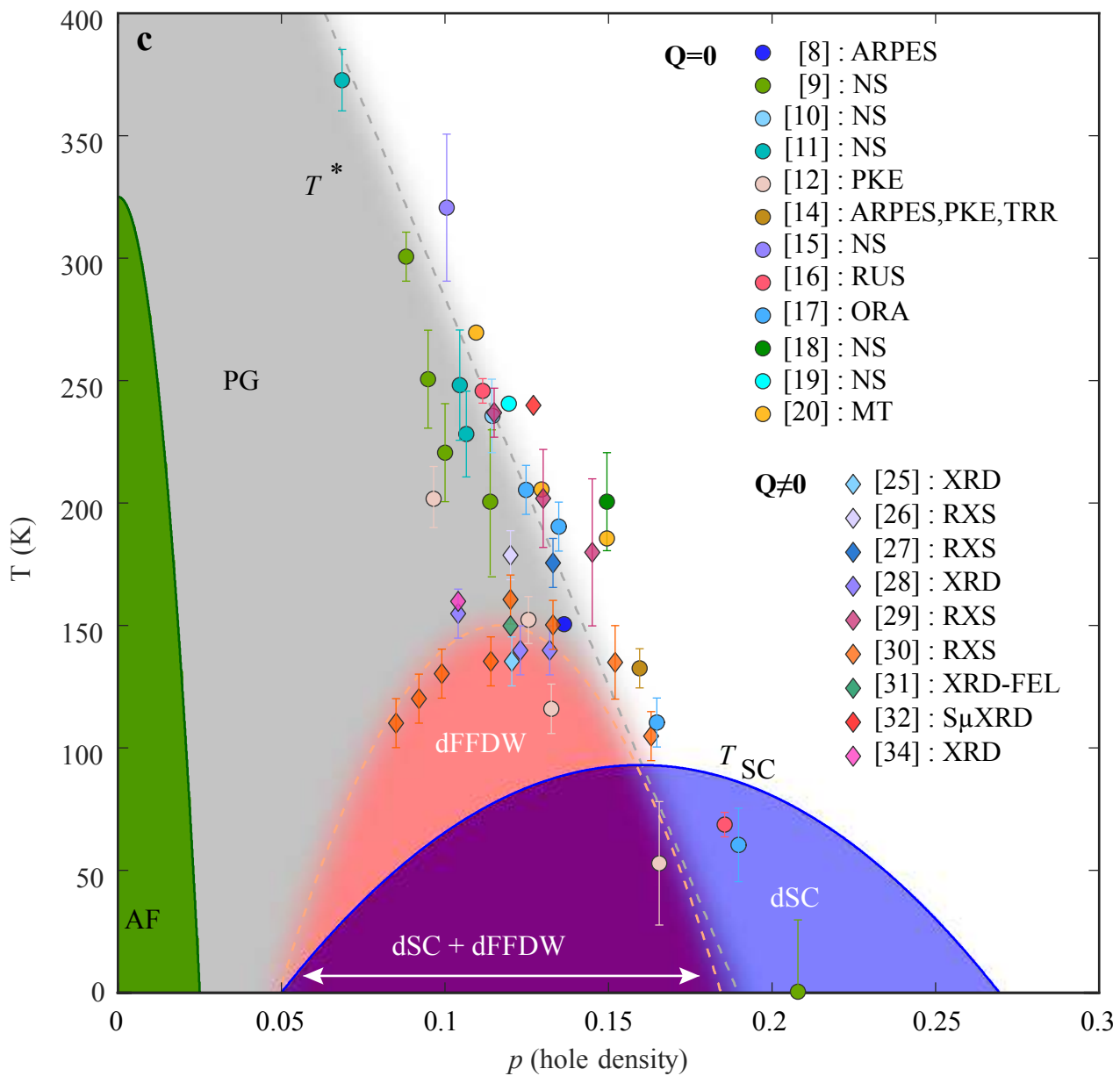

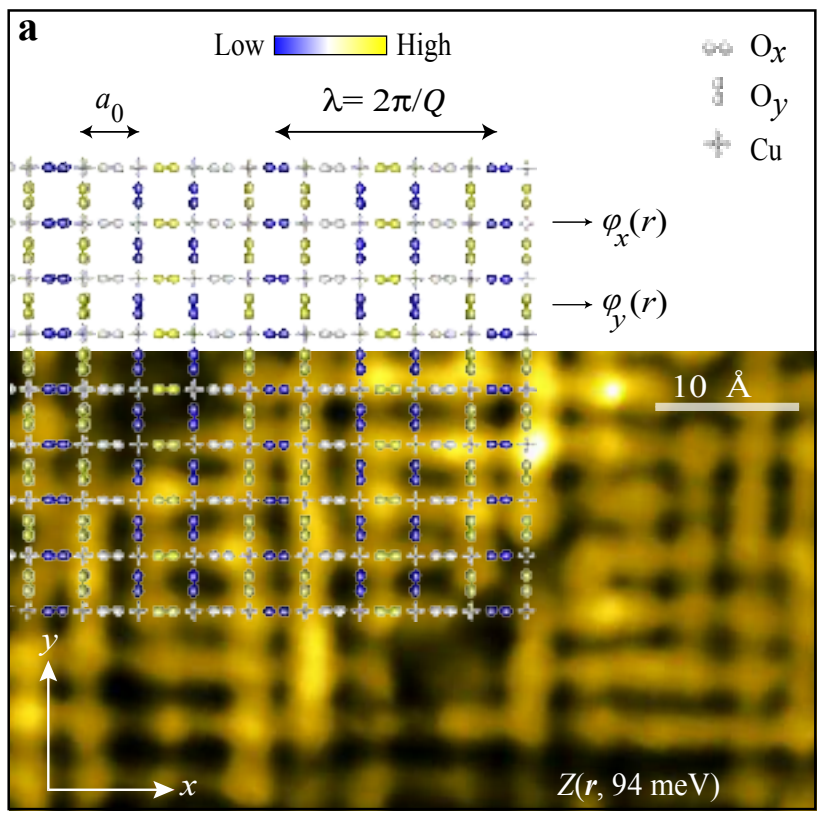

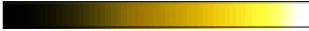

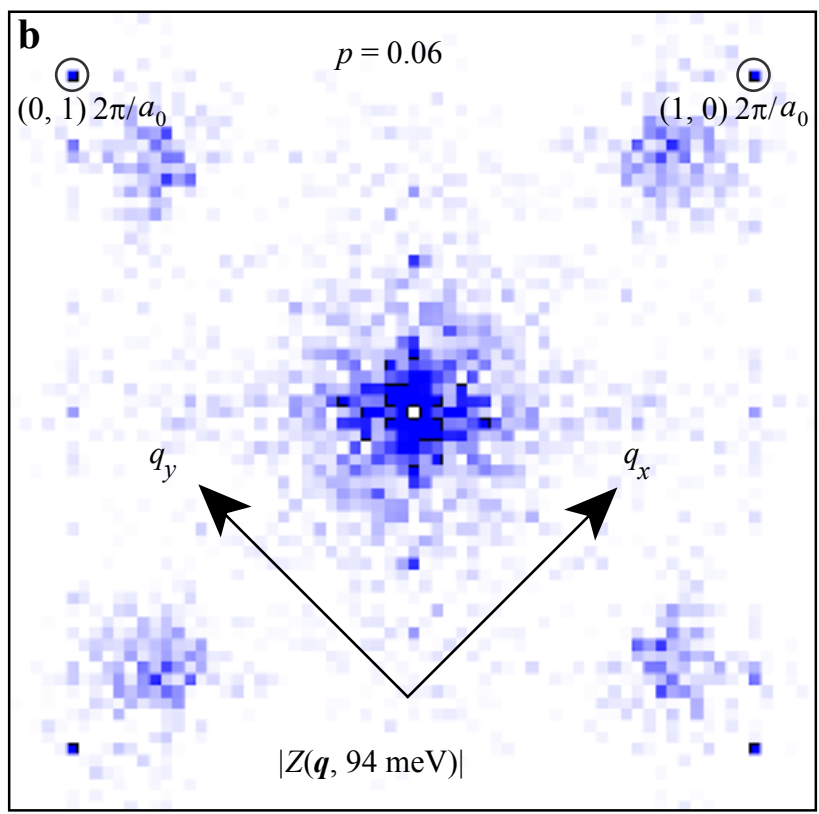

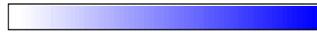

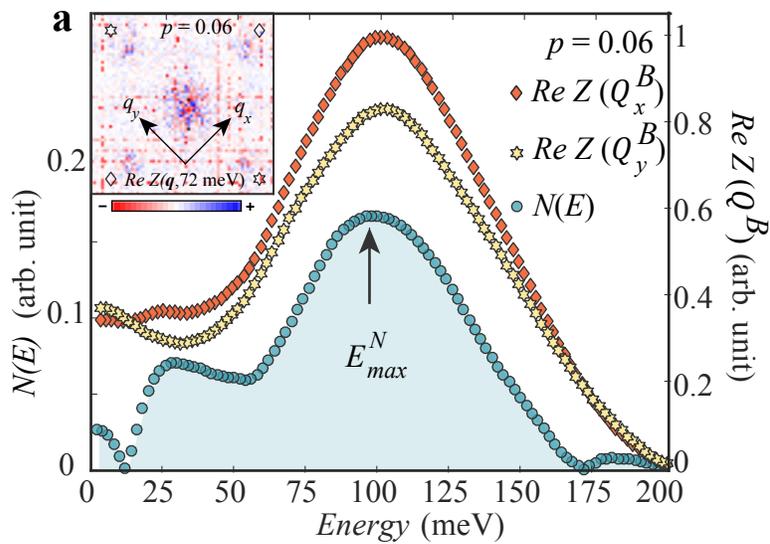
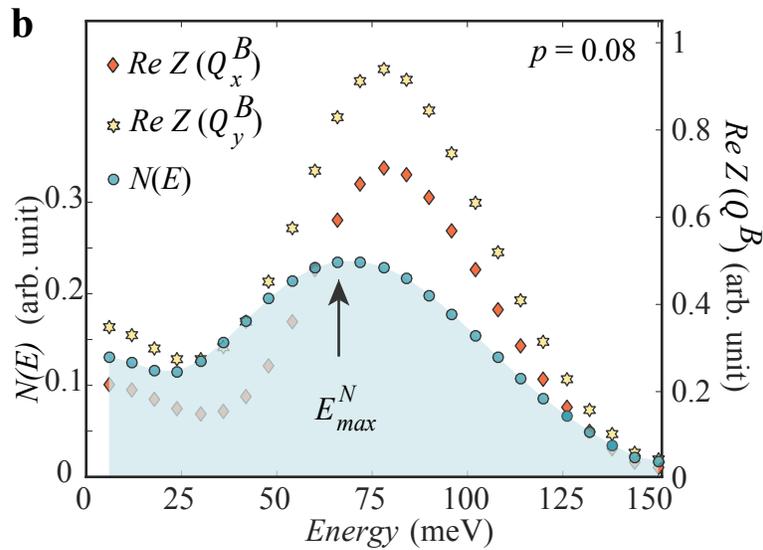
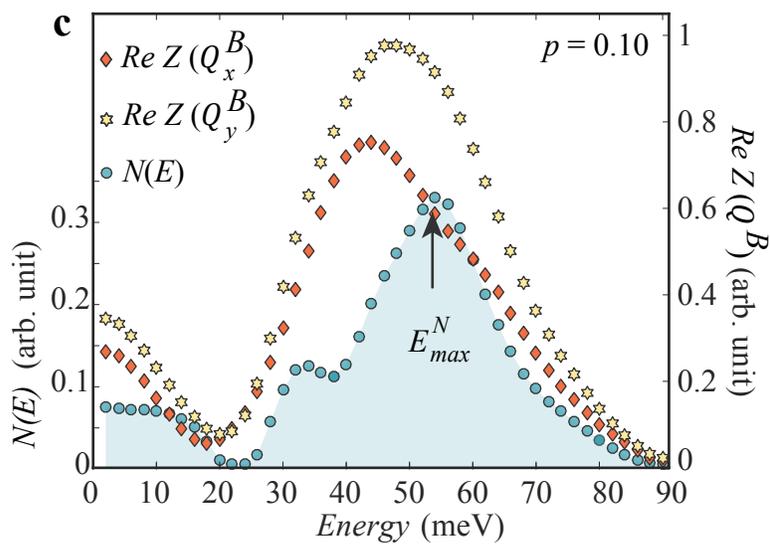
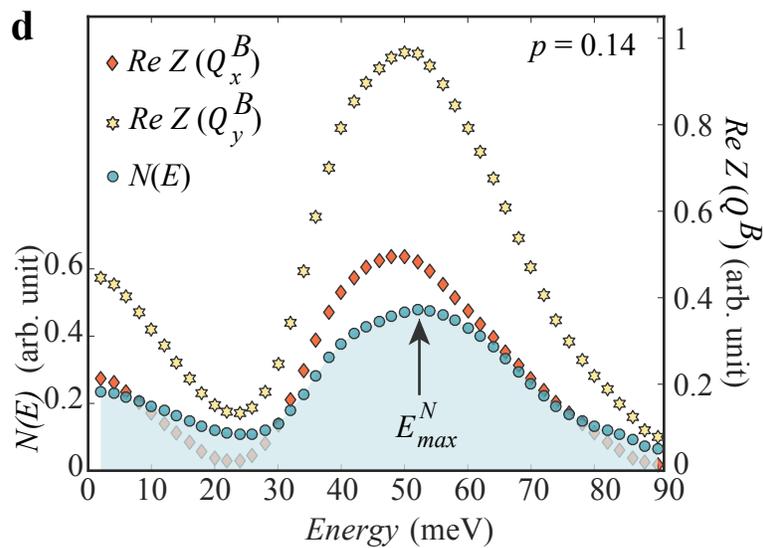
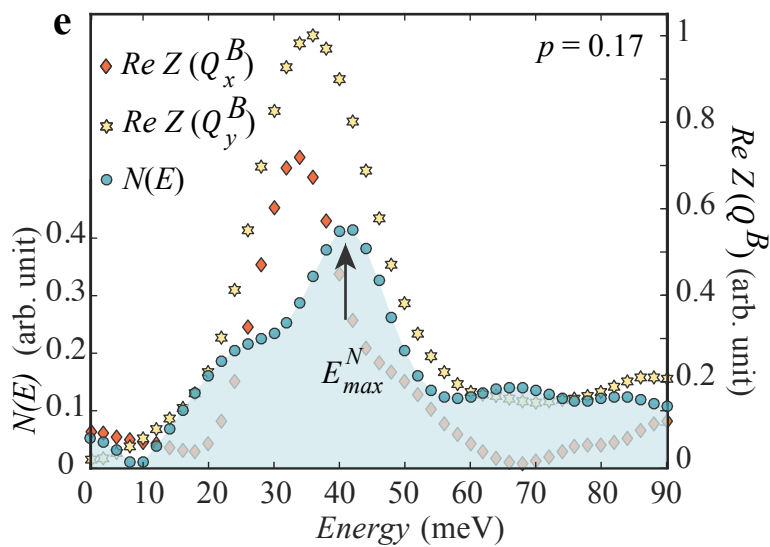
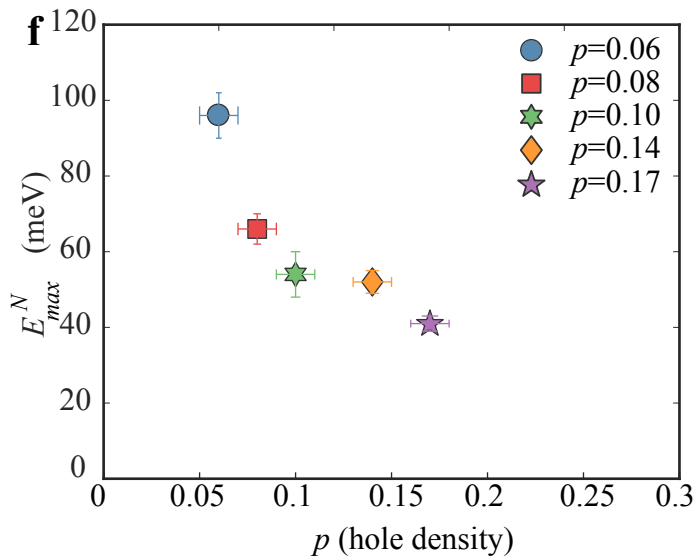

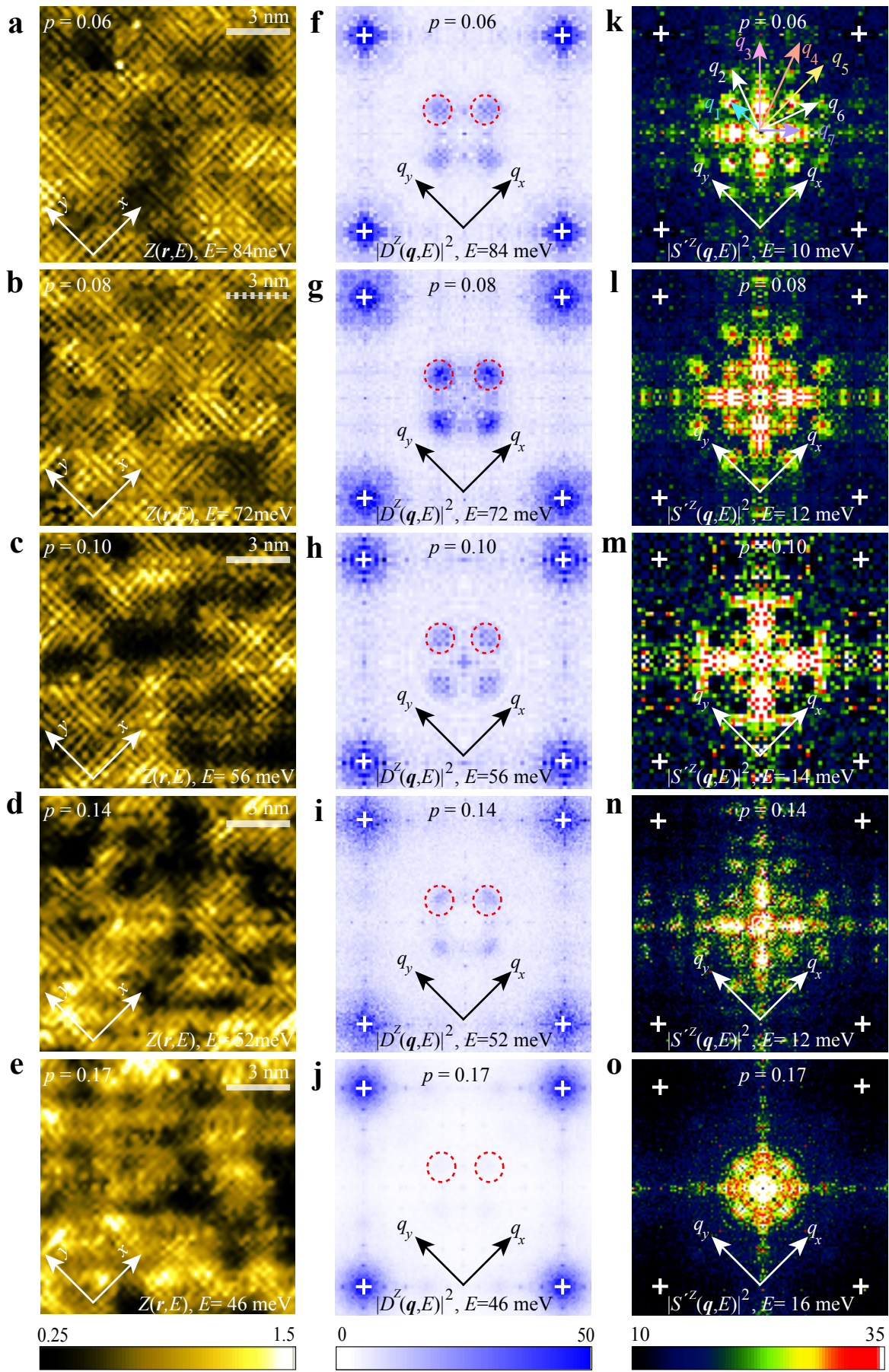

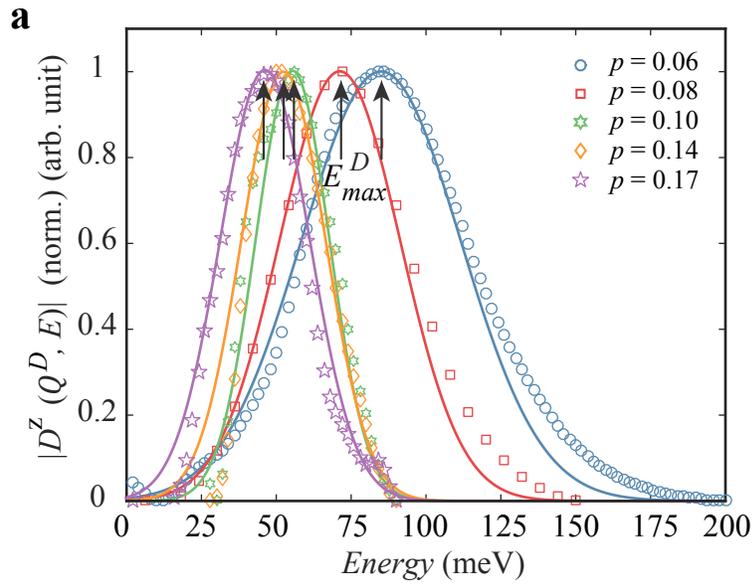

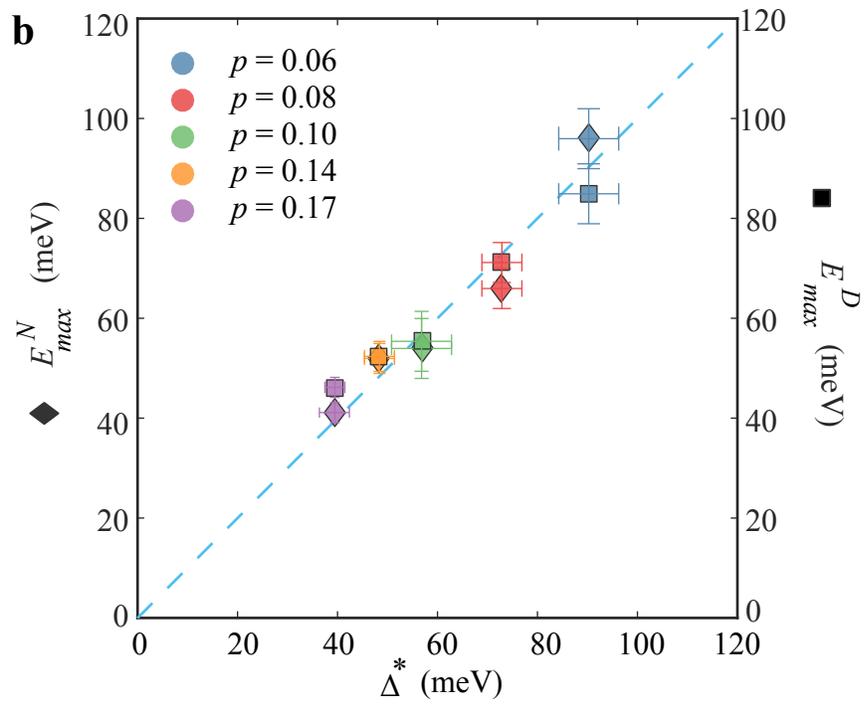

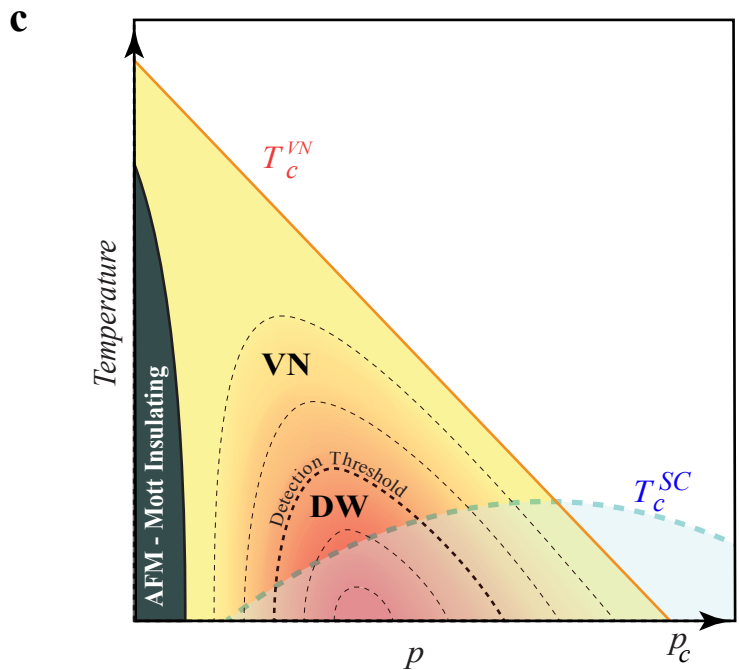

## *Supporting Information for*

## Evidence for a Vestigial Nematic State in the Cuprate Pseudogap Phase

Sourin Mukhopadhyay, Rahul Sharma, Chung Koo Kim, Stephen D. Edkins Mohammad H. Hamidian, Hiroshi Eisaki, Shin-ichi Uchida, Eun-Ah Kim, Michael J. Lawler, Andrew P. Mackenzie, J. C. Séamus Davis and Kazuhiro Fujita

### 1. Bragg amplitude decomposition of the CuO$_2$ NE States

By measuring characteristics of the Bragg peaks in the Fourier transform of an electronic structure image, one is considering the phenomena that occur with the periodicity of the lattice and qualify any C$_4$ symmetry breaking at $\boldsymbol{Q}$=0. For the CuO$_2$ plane, the contributions in the Bragg amplitudes from Cu and O$_x$ and O$_y$ sites in a $Z(\boldsymbol{r})$ image whose Fourier Transform is $\tilde{Z}(\boldsymbol{q})$, are as follows:

$$\tilde{Z}(\boldsymbol{Q}_x^B) = \bar{Z}_{Cu} - \bar{Z}_{O_x} + \bar{Z}_{O_y} \qquad [S1.1]$$

$$\tilde{Z}(\boldsymbol{Q}_y^B) = \bar{Z}_{Cu} + \bar{Z}_{O_x} - \bar{Z}_{O_y} \qquad [S1.2]$$

Here $\bar{Z}_i$ denotes average of $Z(\boldsymbol{r})$ at the atomic site $i$ in the plane. Setting the origin at a Cu atom within the Lawler-Fujita algorithm[1], only the real part of two-dimensional complex Fourier Transform $\tilde{Z}(\boldsymbol{Q})$ enters the definition for $\boldsymbol{Q}$=0 rotational symmetry breaking phenomena[3]. As in eq.1 of the manuscript, one may then define

$$N^Z(E) \equiv Re\tilde{Z}(\boldsymbol{Q}_y^B, E) - Re\tilde{Z}(\boldsymbol{Q}_x^B, E) = 2(\bar{Z}_{O_y} - \bar{Z}_{O_x}) \qquad [S1.3]$$

where $\boldsymbol{Q}_y^B$ and $\boldsymbol{Q}_x^B$ are the Bragg peaks of the CuO$_2$ lattice. Thus, the difference in Bragg amplitudes in eqn. S1.3 is a measure of intra unit-cell-electronic anisotropy about the Cu site, and between the O$_x$ and O$_y$ sites[1,2,3,4]. By contrast, the broad satellite peaks surrounding the Bragg peaks (Fig. 4) contain the information about the local lattice-translation symmetry breaking DW state[1,2].

A finite $N^Z(E)$ might be generated spuriously by a non-rotationally-symmetric (anisotropic) STM tip. For our studies, this is avoided by requiring that topographic images $T(\boldsymbol{r})$, taken with the identical tip used to measure the $Z(r, E)$ data, yield $N^T \equiv Re\tilde{T}(\boldsymbol{Q}_y^B) - Re\tilde{T}(\boldsymbol{Q}_x^B) = 0$. When this $N^T = 0$ condition is achieved, as shown for example in supplementary figures S1A and S1C, it means that the tip detects the crystallographic arrangement of the surface atoms with no rotational symmetry breaking, a situation which cannot occur if the tip geometry is itself anisotropic[1,3,4].

More importantly, one can demonstrate directly from the sub-unit-cell $r$-space imaging of the electronic structure, e.g. supplementary figure S1B, that the source of the inequivalence of Bragg amplitudes is definitely not a spurious broken rotational symmetry of the tip, whose effect must be the same inside every unit cell. Indeed, as shown for example in supplementary figure S1B, the measured electronic structure inside each $CuO_2$ unit cell rotates completely through 90-degerees every 4 unit cells along the modulation direction of the DW. Such a phenomenon is entirely contradictory, at the most elementary level, to expected effects of a spurious tip anisotropy. Moreover, the real part of the Fourier transform of the electronic structure image (figure S1B) exhibits two Bragg amplitudes at $Q_x^B$ and $Q_y^B$ that differ by ~100%, as shown in supplementary figure 1D. We have shown directly, by examining the internal structure of every unit cell individually[3,4], that this finite $N^Z(E)$ effect derives from the imbalance between populations of the two orthogonal arrangements of anisotropic electronic structure at the $O_x$ and $O_y$ site inside each unit cell.

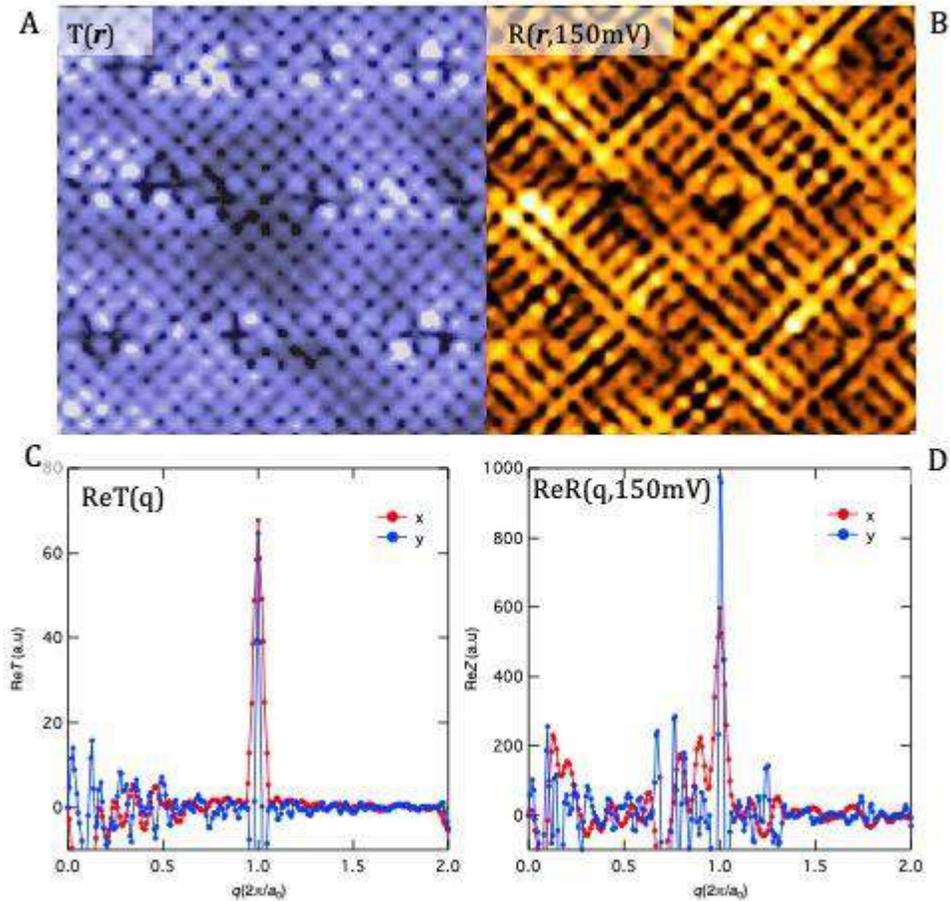

Figure S1A. Topographic image of underdoped $Bi_2Sr_2CaCu_2O_8$; B. Sub-unit-cell resolved electronic structure image $R(r, E = 150meV) \equiv I(r, +150meV)/I(r, -150meV)$ measured simultaneously with A; C. Comparison of $Re\tilde{T}(Q_y^B)$ and $Re\tilde{T}(Q_x^B)$ derived from A, showing that $N^T = 0$ for this experiment; D. Comparison of $Re\tilde{R}(Q_y^B)$ and $ReR(Q_x^B)$ from B showing that $N^R > 0$.

## 2. Form-factor symmetry decomposition of CuO₂ DW States

For the CuO₂ plane, one can think of the angular form factor organization of electronic density waves in terms of $\boldsymbol{Q} = 0$ states whose point group symmetry is well defined. The $\boldsymbol{Q} = 0$ s-wave symmetry has a density

$$D(\boldsymbol{r_{Cu}}) = A_S, \qquad D(\boldsymbol{r_{O_x}}) = 0, D(\boldsymbol{r_{O_y}}) = 0 \qquad [S2.1]$$

the $\boldsymbol{Q} = 0$ s'-wave symmetry has density

$$D(\boldsymbol{r_{Cu}}) = 0, \qquad D(\boldsymbol{r_{O_x}}) = A_{S'}, D(\boldsymbol{r_{O_y}}) = A_{S'} \qquad [S2.2]$$

and the $\boldsymbol{Q} = 0$ d-wave symmetry has density

$$D(\boldsymbol{r_{Cu}}) = 0, \qquad D(\boldsymbol{r_{O_x}}) = A_D, D(\boldsymbol{r_{O_y}}) = -A_D \qquad [S2.3]$$

Modulating these states periodically in a DW, one obtains

$$D_S(\boldsymbol{r}) = \begin{cases} A_S \cos(\boldsymbol{Q}\cdot\boldsymbol{r} + \phi_S), & \boldsymbol{r} = \boldsymbol{r_{Cu}}, \\ 0, & \boldsymbol{r} = \boldsymbol{r_{O_x}}, \\ 0, & \boldsymbol{r} = \boldsymbol{r_{O_y}}, \end{cases} \quad D_{S'}(\boldsymbol{r}) = \begin{cases} 0, & \boldsymbol{r} = \boldsymbol{r_{Cu}}, \\ A_{S'} \cos(\boldsymbol{Q}\cdot\boldsymbol{r} + \phi_{S'}), & \boldsymbol{r} = \boldsymbol{r_{O_x}}, \\ A_{S'} \cos(\boldsymbol{Q}\cdot\boldsymbol{r} + \phi_{S'}), & \boldsymbol{r} = \boldsymbol{r_{O_y}}, \end{cases}$$

$$D_D(\boldsymbol{r}) = \begin{cases} 0, & \boldsymbol{r} = \boldsymbol{r_{Cu}}, \\ A_D \cos(\boldsymbol{Q}\cdot\boldsymbol{r} + \phi_D), & \boldsymbol{r} = \boldsymbol{r_{O_x}}, \\ -A_D \cos(\boldsymbol{Q}\cdot\boldsymbol{r} + \phi_D), & \boldsymbol{r} = \boldsymbol{r_{O_y}}, \end{cases} \qquad [S2.4]$$

The s-wave form factor is a wave purely on the copper atoms with no weight on the oxygen atoms, while the s'-wave and d-wave form factors involve purely the oxygen sites. Therefore, one analyses the segregated oxygen sub-lattice images $O_{x,y}(\boldsymbol{r})$. In terms of the segregated sub-lattices, a d-form factor DW is one for which the DW on the $O_x$ sites is in anti-phase with that on the $O_y$ sites. Hence the peaks at $\pm\boldsymbol{Q_x}$ and $\pm\boldsymbol{Q_y}$ present in both $\tilde{O}_x(\boldsymbol{q})$ and $\tilde{O}_y(\boldsymbol{q})$ must cancel exactly in $\tilde{O}_x(\boldsymbol{q}) + \tilde{O}_y(\boldsymbol{q})$ and be enhanced in $\tilde{O}_x(\boldsymbol{q}) - \tilde{O}_y(\boldsymbol{q})$ as can be seen from eq. S3.4. This is why Eqn. 2a in the manuscript is used to calculate the amplitude of the d-symmetry form factor DW.

Overall, it is important to reemphasize that the DW state described here and detected by experiments is then a unidirectional modulation with $|\boldsymbol{Q}| = 2\pi/4a_0$, of the intra-unit-cell rotation symmetry breaking described by $N^Z(E)$ in eqn S1.1 (see Refs. 3,4).

### 3. Maximum in the density of states D(E) at the gap edge of CDW systems:

Consider a weakly coupled 1-D electron-lattice system under mean field theory. Under independent electron, harmonic and adiabatic approximations, such a system can be described by a Fröhlich Hamiltonian:

$$H_{el-ph} = \sum_{k} \epsilon_k a_k^+ a_k + \sum_{q} \hbar\omega_q b_q^+ b_q + \frac{1}{\sqrt{N}} \sum_{k,q} g_q\, a_{k+q}^+ a_k (b_{-q}^+ + b_q) \qquad [S3.1]$$

where $\epsilon_k, \hbar\omega_q$ and $g_q$ denote energy of electronic state **k**, energy of phonon mode **q** and electron-phonon coupling constant (assumed independent of **k**). $a_k^+$ ($a_k$) creates (annihilates) an electron state **k** and $b_q^+$ ($b_q$) creates (annihilates) a phonon state **q**. The mean field charge density wave CDW order parameter is then

$$\Delta = g_{2k_F} \langle b_{-2k_F}^+ + b_{2k_F} \rangle e^{i 2 k_F x} \qquad [S3.2]$$

This mean field can be used to diagonalize the electronic Hamiltonian and results in energy eigenmodes $E_k$ separated by a gap of 2Δ:

$$E_k = \begin{cases} -\sqrt{\epsilon_k^2 + \Delta^2}, & \epsilon_k < 0 \\ \sqrt{\epsilon_k^2 + \Delta^2}, & \epsilon_k \geq 0 \end{cases} \qquad [S3.3]$$

The number of electrons is conserved and thus so is the density of states. In that case, $D_{CDW}(E)dE = D_n(\epsilon)d\epsilon$, where $n$ denotes the normal state. Assuming that $\Delta \ll E_F$ since we are interested in in energies around $E_F = 0$, we can take $D_n(\epsilon) \approx D_n(0)$, a constant. This leads to a simple relation for the density of states D(E):

$$\frac{D_{CDW}(E)}{D_n(0)} = \frac{d\epsilon_k}{dE_k} = \begin{cases} 0, & E_k < \Delta \\ \dfrac{E_k}{\sqrt{E_k^2 - \Delta^2}}, & E_k \geq \Delta \end{cases} \qquad [S1.4]$$

Hence $D_{CDW}(E)$ exhibits two sharp maxima at two gap-edge energies $E_{max}^{CDW} = \pm\Delta$.

In a more realistic case for this CDW model when no longer in the weakly coupled 1-D limit, the electronic susceptibility is not logarithmically divergent and hence its peaks are not singular but finite. However, the maxima in $D_{CDW}(E)$ still occur at the two gap edges $E_{max}^{CDW} = +\Delta_+; -\Delta_-$ which are no longer necessarily particle-hole symmetric.